\documentclass[twocolumn,prl,superscriptaddress]{revtex4}

\usepackage{graphicx}
\usepackage{dcolumn}
\usepackage{bm}
\usepackage{multirow}
\usepackage{color}
\usepackage[normalem]{ulem}
\usepackage{amsmath}
\usepackage{gensymb}
\usepackage[colorlinks=true,citecolor=blue,linkcolor=blue, urlcolor=cyan]{hyperref} 

\usepackage{xcolor}
\usepackage[normalem]{ulem}
\usepackage[makeroom]{cancel}

\begin{document}

\title{Spin-valley polarization control in WSe$_2$ monolayers using photochemical doping}

\author{E.~Katsipoulaki}
\affiliation{Institute of Electronic Structure and Laser, Foundation for Research and Technology-Hellas, Heraklion, 70013, Greece}
\affiliation{Department of Physics, University of Crete, Heraklion 70013, Greece}
\author{K.~Mourzidis}
\affiliation{Universite de Toulouse, INSA-CNRS-UPS, LPCNO, 135 Av. Rangueil, 31077 Toulouse, France}
\author{V.~Jindal}
\affiliation{Universite de Toulouse, INSA-CNRS-UPS, LPCNO, 135 Av. Rangueil, 31077 Toulouse, France}
\author{D.~Lagarde}
\affiliation{Universite de Toulouse, INSA-CNRS-UPS, LPCNO, 135 Av. Rangueil, 31077 Toulouse, France}
\author{T.~Taniguchi}
\affiliation{Research Center for Materials Nanoarchitectonics, National Institute for Materials Science,  1-1 Namiki, Tsukuba 305-0044, Japan}
\author{K.~Watanabe}
\affiliation{Research Center for Electronic and Optical Materials, National Institute for Materials Science, 1-1 Namiki, Tsukuba 305-0044, Japan}
\author{G.~Kopidakis}
\affiliation{Institute of Electronic Structure and Laser, Foundation for Research and Technology-Hellas, Heraklion, 70013, Greece}
\affiliation{Department of Materials Science and Engineering, University of Crete, Heraklion 70013, Greece}
\author{X.~Marie}
\email{xavier.marie@insa-toulouse.fr}
\affiliation{Universite de Toulouse, INSA-CNRS-UPS, LPCNO, 135 Av. Rangueil, 31077 Toulouse, France}
\affiliation{Institut Universitaire de France, 75231 Paris, France}
\author{M.M.~Glazov}
\email{glazov@coherent.ioffe.ru}
\affiliation{Ioffe Institute, 194021 St. Petersburg, Russia}
\author{E.~Stratakis}
\email{stratak@iesl.forth.gr}
\affiliation{Institute of Electronic Structure and Laser, Foundation for Research and Technology-Hellas, Heraklion, 70013, Greece}
\author{G.~Kioseoglou}
\email{gnk@materials.uoc.gr}
\affiliation{Institute of Electronic Structure and Laser, Foundation for Research and Technology-Hellas, Heraklion, 70013, Greece}
\affiliation{Department of Materials Science and Engineering, University of Crete, Heraklion 70013, Greece}
\author{I.~Paradisanos}
\email{iparad@iesl.forth.gr}
\affiliation{Institute of Electronic Structure and Laser, Foundation for Research and Technology-Hellas, Heraklion, 70013, Greece}

\begin{abstract}
 \textbf{Abstract.}
 We report on the influence of a photochemical doping method on the spin-valley polarization degree ($P_{c}$) of excitons in WSe$_2$ monolayers. By varying the carrier density and transitioning from an excess of electrons (n-type) to an excess of holes (p-type), we observe a non-monotonic dependence of $P_{c}$ on the doping level. Using controlled, single-shot photochlorination steps, we unveil this non-monotonic behavior, with $P_{c}$ reaching a minimum value of less than 10$\%$ at 78 K near the charge neutrality point, while increasing by a factor of three at a hole density of $5 \times 10^{11} \,\mathrm{cm^{-2}}$. The impact of the doping on $P_{c}$ is explained using a phenomenological model that accounts for various mechanisms influencing exciton polarization dynamics, including exciton-carrier scattering processes and exciton-to-trion conversion rates. Among these, exciton-carrier collisions emerge as the dominant mechanism driving the observed variations in $P_{c}$, while the exciton effective lifetime remains nearly independent of doping. These findings highlight the potential of photochemical methods for investigating valley physics and for effectively tuning the exciton polarization degree in transition metal dichalcogenide monolayers.

\end{abstract}

\maketitle

 \textbf{Introduction.}\\
In monolayer transition metal dichalcogenides (TMDs), the local extrema of electronic bands in momentum space at the $K^+$ and $K^-$ points of the Brillouin zone are associated with a quantum number often termed ``valley pseudospin". This degree of freedom has attracted significant interest due to its potential to serve as an additional information carrier, complementing the electron's charge and spin and offering opportunities for encoding, processing, and storing information in emerging valleytronic applications\cite{schaibley2016valleytronics}. This concept mirrors spintronics, where the spin of electrons and their associated magnetic moment introduces an additional intrinsic degree of freedom\cite{wolf2001spintronics}. The manipulation of valleys is a well-established concept, historically involving inversion layers at silicon/insulator interfaces\cite{sham1978valley}, as well as strain\cite{thompson200490} and magnetic fields in 2D electron-gas systems\cite{shkolnikov2002valley}. The primary aim of valleytronics is to develop systems that can provide clear advantages in terms of processing speed, energy efficiency, and information storage, thereby complementing and surpassing contemporary semiconductor technologies based on charge and spin.\\ 
\indent The origin of the ``valley pseudospin" in TMDs lies primarily in the symmetry of electronic states near the $K$-points, complying with time inversion but requiring a broken inversion symmetry, a condition met in odd-numbered TMD layers\cite{liu2015electronic}. Specifically, monolayers of TMDs exhibit a direct bandgap at the $K^{+}$ and $K^{-}$ points of the Brillouin zone, where chiral optical selection rules apply. These valleys can be selectively populated by circularly polarized light with opposite helicity, enabling the optical generation and detection of valley polarization (as the spin and valley degrees of freedom are locked, we will use in the following the single term ``valley")\cite{xiao2012coupled}. This phenomenon has been explored not only in linear spectroscopy experiments\cite{cao2012valley,mak2012control}, but also in nonlinear studies\cite{hipolito2017second,ho2020measuring,mouchliadis2021probing,herrmann2023nonlinear}.
\begin{figure*}
\includegraphics[width=0.85\linewidth]{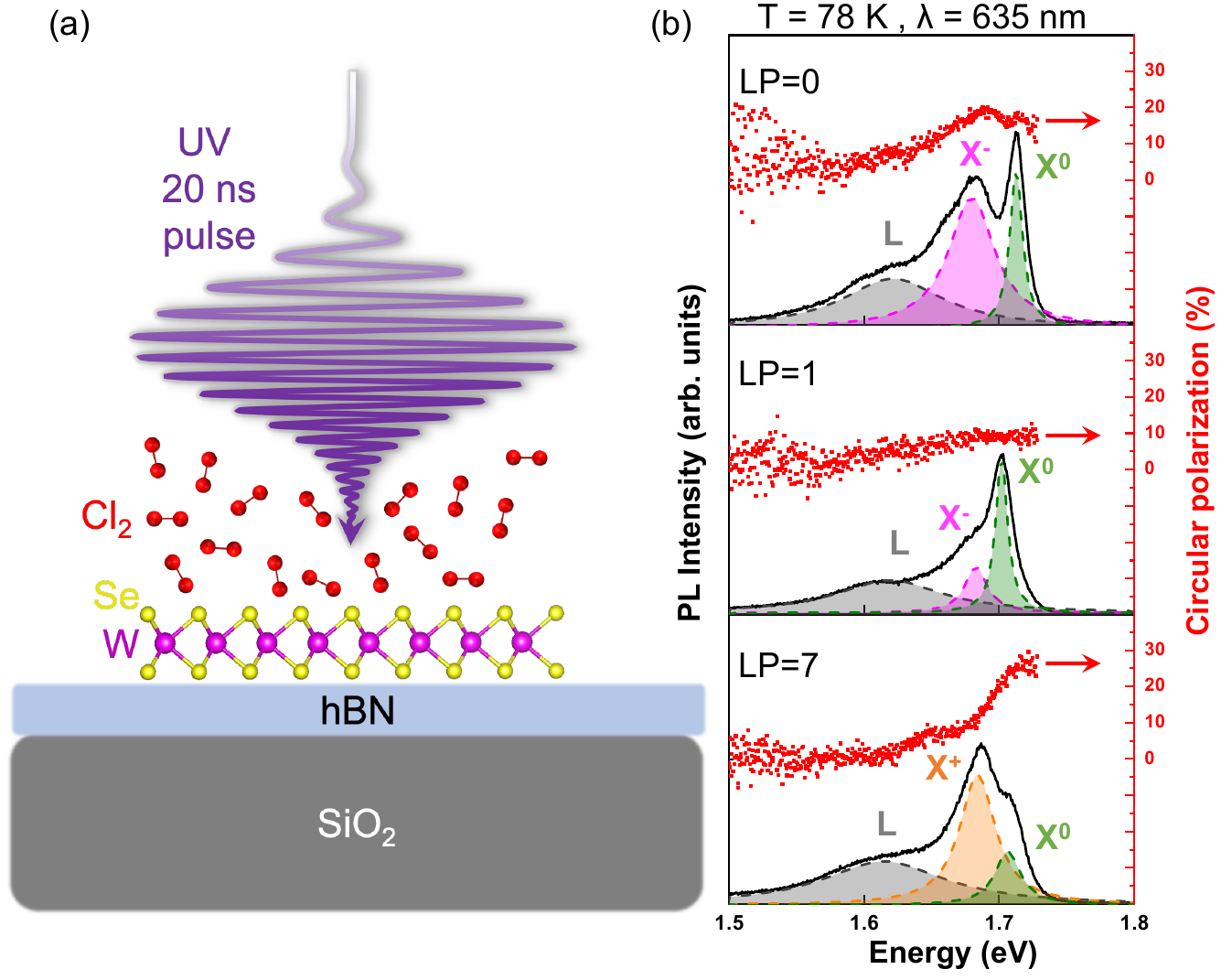}
\caption{\label{fig:fig1} (a) Schematic illustration of the side view of the sample irradiated by a KrF (248~nm,  UV) excimer laser. The metal atoms (W) are represented by magenta spheres, the chalcogen atoms (Se) by yellow spheres, and the Cl atoms by red spheres. The hBN flake is shown in blue, and the SiO$_2$ substrate is depicted in grey. The pulse duration is 20~ns and the repetition rate 1 Hz. (b)~PL spectra of 1L-WSe$_2$/hBN at 78 K with an excitation wavelength of 635 nm for different numbers of UV laser pulses (LP). The corresponding circular polarization is also displayed (red scatter).}
\end{figure*}
Importantly, the optical response of TMD monolayers is primarily governed by excitons, $i.e.$, Coulomb-bound electron-hole pairs\cite{wang2018colloquium}. Photoluminescence (PL) experiments serve as a convenient tool to study the valley degree of freedom, where for one-photon transitions the exciton-photon coupling is determined by the symmetry of the excitonic wave functions\cite{glazov2015spin}. Upon circular excitation, the imbalance between right ($\sigma^{+}$) and left ($\sigma^{-}$) circularly polarized exciton emission intensity is measured to extract the valley polarization degree, $P_{c}$. The value of $P_{c}$ is determined by the effective lifetime, $\tau_{r}$, and the valley polarization lifetime, $\tau_{v}$, through the equation\cite{wang2018colloquium}:
\begin{equation}
P_c=\frac{P_{0}}{1+\frac{\tau_{r}}{\tau_{v}}},
\label{eq1}
\end{equation} 
where $P_{0}$ is the initially generated polarization which reaches unity for close-to-resonance conditions in accordance with the selection rules\cite{glazov2021valley}. Eq. (\ref{eq1}) indicates that $P_{c}$ is significantly influenced by the interplay between $\tau_{r}$ and $\tau_{v}$. In particular, a high $P_{c}$ typically requires a long polarization lifetime as compared to the effective lifetime: $\tau_{r} \ll \tau_v$.\\
\indent Engineering the $P_{c}$ and understanding valley relaxation processes in TMDs pose significant challenges, yet they are pivotal for unlocking the potential of valleytronic applications. The $P_{c}$ in TMD monolayers is conventionally measured under cryogenic temperatures (4 K), magnetic fields, and resonant excitation. Its value is influenced by various factors, including energy detuning\cite{kioseoglou2012valley,wang2015polarization,song2013transport,zeng2012valley,wang2015giant}, temperature\cite{kioseoglou2016optical,hanbicki2016anomalous,paradisanos2020prominent} and strain\cite{zhu2013strain,feierabend2017impact,kourmoulakis2023biaxial,an2023strain}, to name a few examples. Interestingly, while electrostatic doping has recently emerged as an effective approach for modulating valley polarization in gated monolayers ---particularly in the n-type regime \cite{zhang2022prolonging,park2022efficient,feng2019engineering,shinokita2019continuous}--- studies addressing the p-type doping regime remain comparatively scarce, leaving the influence of finite hole density on the $P_c$ less explored.\\
\indent In this work, we overcome the complexities of gated sample fabrication by introducing a straightforward photochemical method to engineer the exciton's circular polarization degree in 1L-WSe$_2$ monolayers. This approach uses single-pulse variation of the carrier density to transition from the n-type to the p-type regime, all without the need for an applied gate voltage\cite{katsipoulaki2023electron}. This method is versatile and can be applied to 1L-WSe$_2$ on arbitrary substrates, offering a significant advantage in terms of simplicity and adaptability. The monolayer doping induced by the photochemical method is first evidenced by the change of the PL spectra exhibiting the recombination of charged excitons.We achieve a three-fold modulation of the exciton polarization degree under non-resonant excitation conditions, further underscoring the effectiveness and robustness of our approach. To support and explain our findings we present a model, concluding that exciton-carrier collisions constitute the most dominant mechanism driving the observed variations in $P_{c}$.

\begin{figure*}
\includegraphics[width=0.85\linewidth]{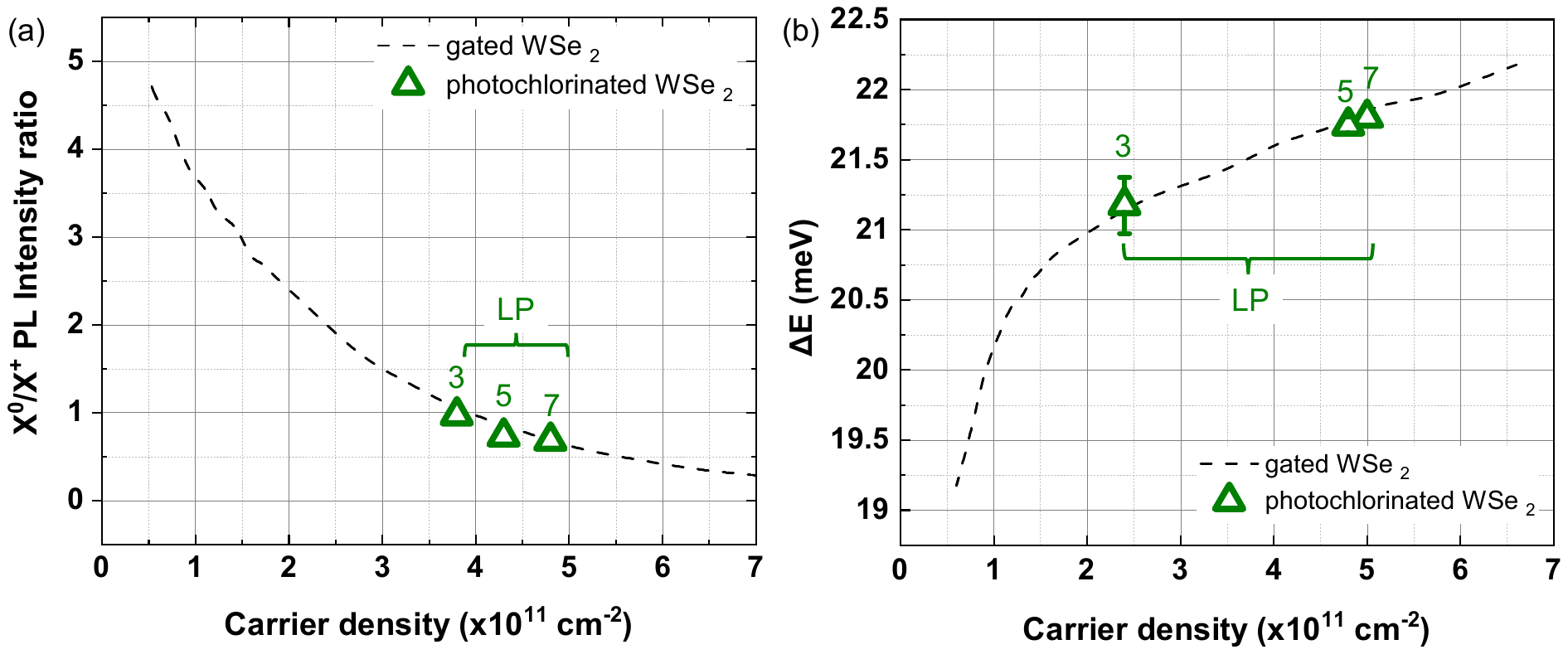}
\caption{\label{fig:fig2} Carrier density calibration of the photochlorination process in the p-type regime. (a) Comparison of the PL intensity ratio of X$^0$ to X$^+$ as a function of hole density between gated (dashed line)\cite{robert2021measurement} and photochlorinated (green triangles) 1L-WSe$_2$. Three, green triangles correspond to different laser pulses, LP=3, LP=5 and LP=7.
(b) Energy difference between X$^0$ and X$^+$ as a function of the hole density for electrostatically gated (dashed line) and photochlorinated (green triangles) monolayer WSe$_2$.}
\end{figure*}

 \textbf{Methods.}\\
The primary focus of this investigation is a 1L-WSe$_2$ on hexagonal boron nitride flakes (hBN) deposited on SiO$_2$/Si substrates (SiO$_2$ is 285 nm-thick). The monolayers are obtained through mechanical exfoliation from a bulk crystal (2D semiconductors) using a Nitto Denko tape (see more details in Sec.~I of supporting information, SI). A schematic representation, illustrating the side view of the heterostructure, is displayed in Fig. \ref{fig:fig1}a. To improve the optical quality of the monolayers, thick ($ \approx $100~nm) hBN flakes are used as supporting substrates\cite{shree2021guide,beret2022exciton,paradisanos2020controlling,ren2023control}. Additionally, detailed information and experimental data concerning reference samples 1L-WSe$_2$ transferred directly to SiO$_2$/Si are available in the SI.  However, the spectral broadening caused by disorder at the SiO$_2$ interface prevents an accurate determination of the peak positions and the intrinsic linewidths. For this reason, we focus on the analysis of samples transferred on hBN for superior optical quality.

For polarization measurements, the sample is excited with circularly polarized light with a photon energy of 1.95~eV (635~nm). Right-handed circularly polarized light ($\sigma^+$) is focused on 1L-WSe$_2$ in a $\mu$-PL (micro-photoluminescence) experimental setup with a combination of a liquid crystal variable retarder (LCVR) and a linear polarizer in the detection path. The emitted PL signal is converted by the LCVR from $\sigma^+$ (positive helicity) and $\sigma^-$ (negative helicity) into the linear basis, and the linear polarizer analyzes the two orthogonal PL components ($I_+$ and $I_-$). The degree of spin-valley polarization is given by
\begin{equation}
P_\text{c} = \frac{I_+ - I_-}{I_+ + I_-}
\label{eq:Pcirc}
\end{equation}
where $I_+$ ($I_-$) represents the intensity of the $\sigma^+$ ($\sigma^-$) components\cite{cao2012valley,xiao2012coupled}, associated with the population imbalance of excited carriers within one of the two inequivalent $K$ valleys.

The photochlorination process involves the irradiation of 1L-WSe$_2$ using a pulsed KrF excimer UV laser at 248~nm, directed through a set of mirrors. The pulse duration is 20~ns, the repetition rate is 1~Hz, and the laser beam fluence is 10~mJ/cm$^2$. The monolayer is positioned inside a vacuum chamber, first pumped down to 10~Torr and subsequently exposed to Cl$_2$ gas with a pressure of 120~Torr. Each photochlorination step involves exposing the monolayer to a single UV pulse in the presence of Cl$_2$, followed by characterization using optical spectroscopy. This process is repeated, with the accumulated laser pulses (LP) changing the carrier density in the material due to adsorption of Cl$_2$ species on the 1L-WSe$_2$ chalcogen vacancy sites. The entire method is carried out at room temperature; for more information on the adsorption mechanism, see\cite{demeridou2018spatially,katsipoulaki2023electron}. An optical microscopy image along with the corresponding Raman spectrum of the sample is shown in Sec.~I,II of the SI. The monolayer thickness is verified by Raman spectroscopy with a prominent peak observed at approximately 250~cm$^{-1}$ corresponding to the two degenerate first-order Raman modes $E'$ and $A'_1$ of 1L-WSe$_2$ along with their 11 cm$^{-1}$ difference from the second order 2LA(M) mode\cite{sahin2013anomalous,hanbicki2015measurement,del2014excited}. 

\begin{figure*}
\includegraphics[width=1\linewidth]{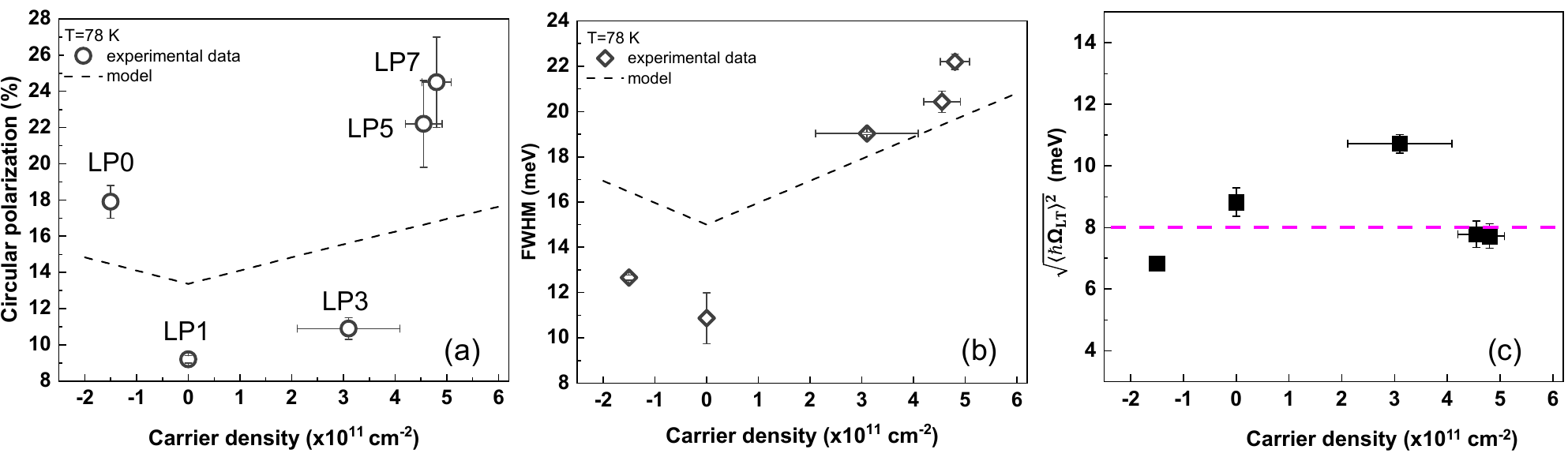}
\caption{\label{fig:fig3} (a)~Circular polarization of X$^0$ as a function of carrier density at 78 K, with circularly polarized laser excitation of 635 nm.  LP indicates the number of UV photochlorination pulses. (b)~FWHM of X$^0$ as a function of carrier density at 78 K.  The black dashed lines in (a) and (b) represent model calculations.  (c)  Root mean square of the longitudinal-transverse splitting for different carrier densities. Pink dashed line corresponds to the average value.}
\end{figure*}

 \textbf{Results.}\\
In Fig. \ref{fig:fig1}b, PL spectra of 1L-WSe$_2$ for the pristine monolayer (LP=0), as well as two indicative photochlorination steps (LP=1 and LP=7) are presented. The circular polarization of the complete PL spectral range is also demonstrated with red points,  extracted by Eq.~(\ref{eq:Pcirc}). In the pristine sample (LP=0), the PL spectrum is predominantly characterized by X$^0$ at 1.712~eV and a lower energy peak at 1.683~eV. The latter is attributed to negatively charged trions, X$^-$ (the sample is naturally electron-doped) due to its characteristic energy difference ($\Delta E$) from X$^0$ of approximately 30~meV\cite{yang2022relaxation,courtade2017charged}. Other emission channels, including those involving localized states and phonon replicas, result in a broad feature (L) near 1.6 eV. Following the initial irradiation step (LP=1), a reduction of X$^-$ intensity is observed, accompanied by a noticeable redshift of X$^0$ compared to its initial energy position. This observation is in agreement with a strong reduction of the residual electron density of the pristine sample due to the partial passivation of defect sites via adsorption of Cl$_2$ molecules\cite{katsipoulaki2023electron}. Upon increasing the laser pulses to LP=7, a new peak emerges $\approx$20 meV below X$^0$, attributed to positively charged trions X$^+$\cite{he2020valley}. Notably, the circular polarization degree at the energy of X$^0$ exhibits a significant enhancement at LP=7, reaching $\approx$  25$\%$, compared to just $\approx$  9$\%$ at LP=1. We note that neutral biexcitons (DX$^0$) also lie close to X$^0$ with a binding energy of 15~meV\cite{yang2022relaxation}. However, the contribution from DX$^0$ in our optical spectra is excluded by observing a linear power dependence using the same excitation wavelength (see Sec.~III in SI)\cite{ye2014probing,kim1994thermodynamics,barbone2018charge,paradisanos2017room}.

The intensity of X$^+$ raises with the number of irradiation laser pulses, from LP=3 to LP=7, reflecting an increase in the hole density of the material. This effect saturates after LP=7, likely due to the complete passivation of defect sites through Cl$_2$ adsorption. A similar behavior is observed in the reference 1L-WSe$_2$/SiO$_2$/Si sample, as shown in Sec.~IV of the SI. Determining the precise hole density for the irradiation steps where X$^+$ appears is crucial for understanding the exciton polarization, as discussed later. A reliable way to convert the photochlorination pulses into carrier density is to compare the optical response between the photochlorinated 1L-WSe$_2$ and a calibrated, electrostatically gated 1L-WSe$_2$\cite{robert2021measurement}. We employ two approaches to estimate the hole density after LP=3, LP=5 and LP=7: one is based on the intensity ratio (Fig. \ref{fig:fig2}a) and the other one on the energy difference (Fig. \ref{fig:fig2}b) between X$^0$ and X$^+$ in the p-type regime. A comparison of the X$^0$/X$^+$ PL intensity ratio for the electrostatically gated 1L-WSe$_2$ (black dashed line in Fig. \ref{fig:fig2}a) and the photochlorinated sample (green triangles in Fig. \ref{fig:fig2}a) indicates an increase in hole density from approximately 4$\times$10$^{11}$ cm$^{-2}$ to 5$\times$10$^{11}$ cm$^{-2}$ as the irradiation progresses from LP=3 to LP=7\cite{robert2021measurement}. Using a similar approach, the residual electron density of the pristine sample (LP=0) is estimated to be 1.5$\times$10$^{11}$ cm$^{-2}$, with the sample approaching the carrier neutrality point after a single irradiation pulse (LP=1), not shown in Fig. \ref{fig:fig2}a. The energy difference ($\Delta E$) between X$^0$ and X$^+$ is also sensitive to the hole density due to renormalization of the exciton and trion energies, as well as the band gap and binding energy of excitons and trions in the presence of a Fermi sea of holes (Fig. \ref{fig:fig2}b)\cite{glazov2020optical}. Based on $\Delta E$, the estimated hole density ranges from approximately $2.5 \times 10^{11} \mathrm{cm}^{-2}$ to $5 \times 10^{11} \mathrm{cm}^{-2}$ when comparing the electrostatically gated 1L-WSe$_2$ (dashed line) to the photochlorinated sample (green triangles), see Fig. \ref{fig:fig2}b. The small divergence at LP=3 between Fig. \ref{fig:fig2}a and Fig. \ref{fig:fig2}b is attributed to the inhomogeneous broadening of the sample, which affects the fitting process used to determine the precise positions and intensities of X$^0$ and X$^+$. To statistically estimate the carrier density at different photochlorination steps, both methods (intensity ratio and energy difference) are averaged, with the standard deviation included in Fig.~\ref{fig:fig3}. Overall, we conclude that as the photochlorination progresses from LP=0 to LP=7, the carrier density of 1L-WSe$_2$ transitions from $1.5 \times 10^{11} \mathrm{cm}^{-2}$ electrons (n-type) to $5 \times 10^{11} \mathrm{cm}^{-2}$ holes (p-type).

We now analyze the influence of carrier density on the PL circular polarization degree, $P_\text{c}$, of X$^0$. Figure \ref{fig:fig3}a presents the variation of $P_\text{c}$ across different photochlorination steps, ranging from LP=0 to LP=7. A non-monotonic behavior is observed: $P_\text{c}$ starts at 18$\%$ when the sample is doped with $1.5 \times 10^{11} \mathrm{cm}^{-2}$ electrons, decreases to 9$\%$ near the neutrality point, and then gradually rises to 25$\%$ at $5 \times 10^{11} \mathrm{cm}^{-2}$ holes. This remarkable modulation of $\approx 3$ highlights the significant impact of carrier density on the spin-valley relaxation of excitons in the material. To gain further insight into the interaction between excitons and carriers, we examine various factors that may influence exciton relaxation processes. 

\begin{figure*}
\includegraphics[width=0.85\linewidth]{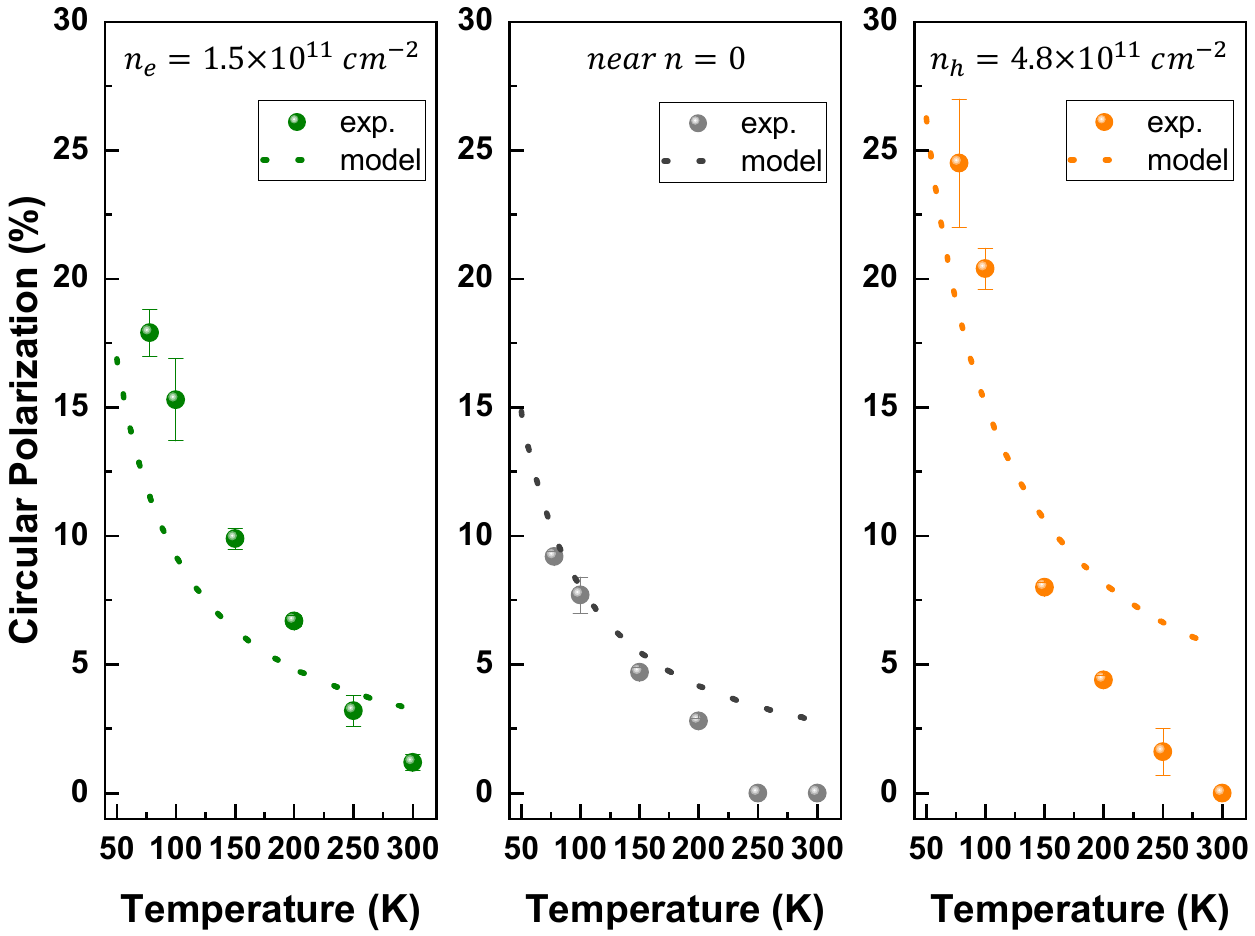}
\caption{\label{fig:fig4} Circular polarization of neutral excitons as a function of temperature for different carrier densities, using a 635 nm excitation wavelength. Experimental values are presented by spheres while the model is shown with dotted curves.}
\end{figure*}

We recall that the main mechanism of the valley relaxation of bright excitons in TMD monolayers is related to the long-range exchange interaction between electron and hole in the exciton~\cite{maialle93,ivchenko05a,glazov2014exciton}. The exchange interaction produces a longitudinal-transverse splitting of excitons $\hbar\Omega_{\rm LT}$ and acts as an effective magnetic field that causes exciton pseudospin precession while the scattering processes randomize the precession and result in the valley polarization decay with the rate~\cite{d1971spin,maialle93,glazov2015spin}
\begin{equation}
\label{tau:v}
\frac{1}{\tau_v} = \langle \Omega_{\rm LT}^2 \tau_{sc}\rangle,
\end{equation} 
where $\tau_{sc}$ is the scattering time, $\Omega_{\rm LT}\tau_{sc} \ll 1$ is assumed, and angular brackets denote the ensemble average. The doping, generally, affects both $\Omega_{\rm LT}$ and $\tau_{sc}$ in Eq.~\eqref{tau:v}.
Screening of the long-range exchange interaction in the presence of carriers (i.e., reduction of $\Omega_{\rm LT}$) has been proposed as the main mechanism to prolong the valley relaxation time\cite{konabe2016screening,shinokita2019continuous,feng2019engineering}. However, the carrier's Fermi energy given by (under an assumption of population only in the bottom spin subband in each valley):
\begin{equation}
E_F = \frac{\pi \hbar^2 n}{m^*},
\label{eq:Efermi}
\end{equation}
where $n$ is the carrier density and $m^*$ the effective mass, varies from $\approx$0.6 meV for $1.5 \times 10^{11} \mathrm{cm}^{-2}$ electrons to $\approx$3 meV for $5 \times 10^{11} \mathrm{cm}^{-2}$ holes in the present work.  For the exchange interaction, the screening should be taken at the frequency $\omega \approx \omega_0$, where $\omega_0$ is the exciton resonance frequency (in 1L-WSe$_2$, $\hbar\omega_0 \approx$1.7 eV)~\cite{Kiselev74,glazov2018breakdown}. Consequently, $\hbar\omega_0$ exceeds by far the carrier's $E_F$ for the samples studied here and, hence, at such high frequencies the resident carriers are unable to react to the rapidly varying fields created by the electron-hole interaction making screening inefficient\cite{glazov2018breakdown}.

The exciton-to-trion conversion can influence the relaxation dynamics of excitons, leading to a shorter effective lifetime, $\tau_{r}$, of X$^0$ resulting in higher $P_\text{c}$\cite{zhang2022prolonging}. The estimates of the trion formation rate presented in~\cite{PhysRevB.102.125410} show that variation of $\tau_r$ in a given density range is negligible. These estimates are further supported by experimental studies of trion dynamics at doping densities similar to those explored here, revealing a trion PL rise time on the picosecond scale\cite{beret2023nonlinear}. At the same time, we observe significant variation of the full width at half maximum (FWHM) of X$^0$ on carrier density. Figure~\ref{fig:fig3}b shows that the FWHM(X$^0$) decreases by approximately 1~meV when transitioning from the electron-rich regime to near the neutrality point, and then increases by about 10~meV as the hole density reaches $5 \times 10^{11} \mathrm{cm}^{-2}$ which cannot be ascribed to the trion formation process.  To investigate whether $\tau_r$ contributes to the observed variations in the FWHM, time-resolved photoluminescence (TRPL) measurements were performed using an excitation wavelength of 633~nm to compare the effective lifetime of X$^0$ in intrinsic (LP=1) and p-doped (LP=7) 1L-WSe$_2$ samples (Sec.~V in SI). For both samples, $\tau_r$ is found to be similar, with a value of approximately 1~ps at 80~K, primarily governed by non-radiative decay channels. These results are consistent across various positions on the samples. 

Hence, we conclude that (i) the variation of $P_c$ with doping is unrelated to $\tau_r$ and, (ii) the FWHM is controlled by the scattering processes with FWHM$=\hbar/\tau_{sc}$. Indeed, exciton-carrier scattering is known to be an important homogeneous broadening channel both in classical two-dimensional systems based on quantum wells~\cite{astakhov00} and TMD monolayers~\cite{efimkin2017many,wagner2023diffusion}. According to Eq.~\eqref{tau:v} a reduction of $\tau_{sc}$ with doping slows down exciton valley relaxation resulting in increase of $P_c$ observed in experiments. Analytical model of Ref.~\cite{wagner2023diffusion} results in 
 \begin{equation}
\label{eq:FWHM}
\frac{\hbar}{\tau_{sc}}= \delta + E_F \frac{M_T}{M_X} \frac{\pi}{\ln^2 ( \frac{\delta}{2 E_T}) + \frac{\pi^2}{4}},
\end{equation}
where $\delta$ is the homogeneous contribution to FWHM related to other scattering processes, such as exciton-phonon scattering,  $M_T$ and $M_X$ are the trion and exciton translational masses and $ E_T$ is the trion binding energy. Eq.~\eqref{eq:FWHM} is derived for the degenerate electrons, for non-degenerate charge carriers it is different by a logarithmic factor. Dashed line in Fig.~\ref{fig:fig3}(b) shows the FWHM calculated after Eq.~\eqref{eq:FWHM} in reasonable agreement with the experiment. Some discrepancy with the data results from simplification of the analytical model and possible additional factors that affect FWHM such as local disorder. Using root mean square of the longitudinal-transverse splitting as a fitting parameter and taking $\tau_r=1$~ps, $P_0=100\%$ we obtain the black dashed curve in Fig.~\ref{fig:fig3}(a) for $\langle (\hbar\Omega_{\rm LT})^2\rangle^{1/2}=8$~meV  which is in agreement with experiment. Importantly, we can confirm this value without using any particular theoretical model for the scattering time: Indeed, taking into account the smallness of $P_c$ and combining Eqs.~\eqref{eq1} and \eqref{tau:v}, we obtain $\langle (\hbar\Omega_{\rm LT})^2\rangle = (\hbar \mathrm{FWHM})/(\tau_r P_c)$. Using experimental values of the FWHM for all available densities we get  $\langle (\hbar\Omega_{\rm LT})^2\rangle^{1/2}$ in the range $7$ to $11$~meV, see Fig.~\ref{fig:fig3}c.

We note that the transfer of exciton polarization to resident carriers or transitions from spin-bright to spin-dark states are less important mechanisms in the observed modulation of $P_\text{c}$. 

In Fig.\ref{fig:fig4}, we present the temperature-dependent behavior of circular polarization for specific carrier densities. The corresponding helicity-resolved PL spectra can be found in Sec.~VI of the SI. Stronger variations in the exciton's polarization are observed at low temperatures. As expected, there is a notable decrease in polarization with increasing temperature for each doping level condition of the system. Regardless of doping level, at room temperature ($T=300$~K), the degree of polarization significantly diminishes, nearly approaching zero. It is consistent with the valley depolarization driven by the electron-hole exchange interaction: $\Omega_{\rm LT}$ scales approximately linearly with the wavevector $k$ of exciton resulting in $\tau_v \propto 1/T$~\cite{zhu2014exciton}. Hence, $P_c$ decreases as $T^{-1}$ at the temperature-independent scattering time $\tau_{sc}$ as shown by corresponding dotted curves in Fig.~\ref{fig:fig4} calculated using $\langle (\hbar\Omega_{\rm LT})^2\rangle^{1/2}=8$~meV at $T=78$~K and temperature-independent $\tau_{sc}$ determined from the FWMH at the same temperature. The deviations of the calculated values from the experiment at high-temperatures indicate importance of other scattering processes (including exciton-phonon scattering) and possible non-linearity of the $\Omega_{\rm LT}(k)$ dependence related, e.g., to the dielectric contrast in the structure~\cite{prazdnichnykh2020control}.

 \textbf{Conclusions.}\\
In conclusion, our study demonstrates a strong --by a factor of $3$-- modulation of the circular polarization degree in neutral exciton emission in 1L-WSe$_2$/hBN through a photo-induced chemical doping process. We observe that the circular polarization of X$^0$ increases with carrier density in both n-type and p-type regimes. Between different mechanisms that impact the circular polarization degree, we conclude that exciton-carrier collisions primarily influence the exciton's spin relaxation time ($\tau_v$), while the radiative lifetime ($\tau_r$) remains practically unchanged. These findings highlight the potential of photochemical doping as a useful method for tailoring and fine-tuning the valleytronic properties of 2D materials, with promising applications in optoelectronics and nanophotonics.

 \textbf{Acknowledgements}\\
E.S., G.K., X.M., V.J., D.L. and I.P. acknowledge support by the EU-funded DYNASTY Project, ID: 101079179, under the Horizon Europe framework programme. M.M.G. was supported by RSF project No. 23-12-00142 (theory). D.L. and X.M acknowledge support by the France 2030 government investment plan managed by the French National Research Agency under grant reference PEPR SPIN – SPINMAT ANR-22-EXSP-0007. K.W. and T.T. acknowledge support from the JSPS KAKENHI (Grant Numbers 21H05233 and 23H02052) , the CREST (JPMJCR24A5), JST and World Premier International Research Center Initiative (WPI), MEXT, Japan. E.K. acknowledges support from the Hellenic Foundation for Research and Innovation (HFRI) under the 4th Call for HFRI PhD Fellowships (Fellowship Number: 9231).

 \bibliography{Katsip_biblio}

\begin{thebibliography}{10}
\expandafter\ifx\csname url\endcsname\relax
  \def\url#1{\texttt{#1}}\fi
\expandafter\ifx\csname urlprefix\endcsname\relax\def\urlprefix{URL }\fi
\providecommand{\bibinfo}[2]{#2}
\providecommand{\eprint}[2][]{\url{#2}}

\bibitem{schaibley2016valleytronics}
\bibinfo{author}{Schaibley, J.~R.} \emph{et~al.}
\newblock \bibinfo{title}{Valleytronics in 2d materials}.
\newblock \emph{\bibinfo{journal}{Nature Reviews Materials}}
  \textbf{\bibinfo{volume}{1}}, \bibinfo{pages}{1--15} (\bibinfo{year}{2016}).

\bibitem{wolf2001spintronics}
\bibinfo{author}{Wolf, S.} \emph{et~al.}
\newblock \bibinfo{title}{Spintronics: a spin-based electronics vision for the
  future}.
\newblock \emph{\bibinfo{journal}{science}} \textbf{\bibinfo{volume}{294}},
  \bibinfo{pages}{1488--1495} (\bibinfo{year}{2001}).

\bibitem{sham1978valley}
\bibinfo{author}{Sham, L.}, \bibinfo{author}{Allen~Jr, S.},
  \bibinfo{author}{Kamgar, A.} \& \bibinfo{author}{Tsui, D.}
\newblock \bibinfo{title}{Valley-valley splitting in inversion layers on a
  high-index surface of silicon}.
\newblock \emph{\bibinfo{journal}{Physical Review Letters}}
  \textbf{\bibinfo{volume}{40}}, \bibinfo{pages}{472} (\bibinfo{year}{1978}).

\bibitem{thompson200490}
\bibinfo{author}{Thompson, S.~E.} \emph{et~al.}
\newblock \bibinfo{title}{A 90-nm logic technology featuring strained-silicon}.
\newblock \emph{\bibinfo{journal}{IEEE Transactions on electron devices}}
  \textbf{\bibinfo{volume}{51}}, \bibinfo{pages}{1790--1797}
  (\bibinfo{year}{2004}).

\bibitem{shkolnikov2002valley}
\bibinfo{author}{Shkolnikov, Y.}, \bibinfo{author}{De~Poortere, E.},
  \bibinfo{author}{Tutuc, E.} \& \bibinfo{author}{Shayegan, M.}
\newblock \bibinfo{title}{Valley splitting of alas two-dimensional electrons in
  a perpendicular magnetic field}.
\newblock \emph{\bibinfo{journal}{Physical review letters}}
  \textbf{\bibinfo{volume}{89}}, \bibinfo{pages}{226805}
  (\bibinfo{year}{2002}).

\bibitem{liu2015electronic}
\bibinfo{author}{Liu, G.-B.}, \bibinfo{author}{Xiao, D.}, \bibinfo{author}{Yao,
  Y.}, \bibinfo{author}{Xu, X.} \& \bibinfo{author}{Yao, W.}
\newblock \bibinfo{title}{Electronic structures and theoretical modelling of
  two-dimensional group-vib transition metal dichalcogenides}.
\newblock \emph{\bibinfo{journal}{Chemical Society Reviews}}
  \textbf{\bibinfo{volume}{44}}, \bibinfo{pages}{2643--2663}
  (\bibinfo{year}{2015}).

\bibitem{xiao2012coupled}
\bibinfo{author}{Xiao, D.}, \bibinfo{author}{Liu, G.-B.},
  \bibinfo{author}{Feng, W.}, \bibinfo{author}{Xu, X.} \& \bibinfo{author}{Yao,
  W.}
\newblock \bibinfo{title}{Coupled spin and valley physics in monolayers of
  mos$_2$ and other group-vi dichalcogenides}.
\newblock \emph{\bibinfo{journal}{Physical review letters}}
  \textbf{\bibinfo{volume}{108}}, \bibinfo{pages}{196802}
  (\bibinfo{year}{2012}).

\bibitem{cao2012valley}
\bibinfo{author}{Cao, T.} \emph{et~al.}
\newblock \bibinfo{title}{Valley-selective circular dichroism of monolayer
  molybdenum disulphide}.
\newblock \emph{\bibinfo{journal}{Nature communications}}
  \textbf{\bibinfo{volume}{3}}, \bibinfo{pages}{887} (\bibinfo{year}{2012}).

\bibitem{mak2012control}
\bibinfo{author}{Mak, K.~F.}, \bibinfo{author}{He, K.}, \bibinfo{author}{Shan,
  J.} \& \bibinfo{author}{Heinz, T.~F.}
\newblock \bibinfo{title}{Control of valley polarization in monolayer mos$_2$
  by optical helicity}.
\newblock \emph{\bibinfo{journal}{Nature nanotechnology}}
  \textbf{\bibinfo{volume}{7}}, \bibinfo{pages}{494--498}
  (\bibinfo{year}{2012}).

\bibitem{hipolito2017second}
\bibinfo{author}{Hipolito, F.} \& \bibinfo{author}{Pereira, V.~M.}
\newblock \bibinfo{title}{Second harmonic spectroscopy to optically detect
  valley polarization in 2d materials}.
\newblock \emph{\bibinfo{journal}{2D Materials}} \textbf{\bibinfo{volume}{4}},
  \bibinfo{pages}{021027} (\bibinfo{year}{2017}).

\bibitem{ho2020measuring}
\bibinfo{author}{Ho, Y.~W.} \emph{et~al.}
\newblock \bibinfo{title}{Measuring valley polarization in two-dimensional
  materials with second-harmonic spectroscopy}.
\newblock \emph{\bibinfo{journal}{ACS Photonics}} \textbf{\bibinfo{volume}{7}},
  \bibinfo{pages}{925--931} (\bibinfo{year}{2020}).

\bibitem{mouchliadis2021probing}
\bibinfo{author}{Mouchliadis, L.} \emph{et~al.}
\newblock \bibinfo{title}{Probing valley population imbalance in transition
  metal dichalcogenides via temperature-dependent second harmonic generation
  imaging}.
\newblock \emph{\bibinfo{journal}{npj 2D Materials and Applications}}
  \textbf{\bibinfo{volume}{5}}, \bibinfo{pages}{6} (\bibinfo{year}{2021}).

\bibitem{herrmann2023nonlinear}
\bibinfo{author}{Herrmann, P.} \emph{et~al.}
\newblock \bibinfo{title}{Nonlinear all-optical coherent generation and
  read-out of valleys in atomically thin semiconductors}.
\newblock \emph{\bibinfo{journal}{Small}} \bibinfo{pages}{2301126}
  (\bibinfo{year}{2023}).

\bibitem{wang2018colloquium}
\bibinfo{author}{Wang, G.} \emph{et~al.}
\newblock \bibinfo{title}{Colloquium: Excitons in atomically thin transition
  metal dichalcogenides}.
\newblock \emph{\bibinfo{journal}{Reviews of Modern Physics}}
  \textbf{\bibinfo{volume}{90}}, \bibinfo{pages}{021001}
  (\bibinfo{year}{2018}).

\bibitem{glazov2015spin}
\bibinfo{author}{Glazov, M.~M.} \emph{et~al.}
\newblock \bibinfo{title}{Spin and valley dynamics of excitons in transition
  metal dichalcogenide monolayers}.
\newblock \emph{\bibinfo{journal}{physica status solidi (b)}}
  \textbf{\bibinfo{volume}{252}}, \bibinfo{pages}{2349--2362}
  (\bibinfo{year}{2015}).

\bibitem{glazov2021valley}
\bibinfo{author}{Glazov, M.~M.} \& \bibinfo{author}{Ivchenko, E.~L.}
\newblock \bibinfo{title}{Valley orientation of electrons and excitons in
  atomically thin transition metal dichalcogenide monolayers (brief review)}.
\newblock \emph{\bibinfo{journal}{JETP Letters}}
  \textbf{\bibinfo{volume}{113}}, \bibinfo{pages}{7--17}
  (\bibinfo{year}{2021}).

\bibitem{kioseoglou2012valley}
\bibinfo{author}{Kioseoglou, G.} \emph{et~al.}
\newblock \bibinfo{title}{Valley polarization and intervalley scattering in
  monolayer mos$_2$}.
\newblock \emph{\bibinfo{journal}{Applied Physics Letters}}
  \textbf{\bibinfo{volume}{101}} (\bibinfo{year}{2012}).

\bibitem{wang2015polarization}
\bibinfo{author}{Wang, G.} \emph{et~al.}
\newblock \bibinfo{title}{Polarization and time-resolved photoluminescence
  spectroscopy of excitons in mose$_2$ monolayers}.
\newblock \emph{\bibinfo{journal}{Applied Physics Letters}}
  \textbf{\bibinfo{volume}{106}} (\bibinfo{year}{2015}).

\bibitem{song2013transport}
\bibinfo{author}{Song, Y.} \& \bibinfo{author}{Dery, H.}
\newblock \bibinfo{title}{Transport theory of monolayer transition-metal
  dichalcogenides through symmetry}.
\newblock \emph{\bibinfo{journal}{Physical review letters}}
  \textbf{\bibinfo{volume}{111}}, \bibinfo{pages}{026601}
  (\bibinfo{year}{2013}).

\bibitem{zeng2012valley}
\bibinfo{author}{Zeng, H.}, \bibinfo{author}{Dai, J.}, \bibinfo{author}{Yao,
  W.}, \bibinfo{author}{Xiao, D.} \& \bibinfo{author}{Cui, X.}
\newblock \bibinfo{title}{Valley polarization in mos$_2$ monolayers by optical
  pumping}.
\newblock \emph{\bibinfo{journal}{Nature nanotechnology}}
  \textbf{\bibinfo{volume}{7}}, \bibinfo{pages}{490--493}
  (\bibinfo{year}{2012}).

\bibitem{wang2015giant}
\bibinfo{author}{Wang, G.} \emph{et~al.}
\newblock \bibinfo{title}{Giant enhancement of the optical second-harmonic
  emission of wse$_2$ monolayers by laser excitation at exciton resonances}.
\newblock \emph{\bibinfo{journal}{Physical review letters}}
  \textbf{\bibinfo{volume}{114}}, \bibinfo{pages}{097403}
  (\bibinfo{year}{2015}).

\bibitem{kioseoglou2016optical}
\bibinfo{author}{Kioseoglou, G.}, \bibinfo{author}{Hanbicki, A.~T.},
  \bibinfo{author}{Currie, M.}, \bibinfo{author}{Friedman, A.~L.} \&
  \bibinfo{author}{Jonker, B.~T.}
\newblock \bibinfo{title}{Optical polarization and intervalley scattering in
  single layers of mos$_2$ and mose$_2$}.
\newblock \emph{\bibinfo{journal}{Scientific reports}}
  \textbf{\bibinfo{volume}{6}}, \bibinfo{pages}{25041} (\bibinfo{year}{2016}).

\bibitem{hanbicki2016anomalous}
\bibinfo{author}{Hanbicki, A.~T.} \emph{et~al.}
\newblock \bibinfo{title}{Anomalous temperature-dependent spin-valley
  polarization in monolayer ws$_2$}.
\newblock \emph{\bibinfo{journal}{Scientific reports}}
  \textbf{\bibinfo{volume}{6}}, \bibinfo{pages}{18885} (\bibinfo{year}{2016}).

\bibitem{paradisanos2020prominent}
\bibinfo{author}{Paradisanos, I.} \emph{et~al.}
\newblock \bibinfo{title}{Prominent room temperature valley polarization in
  ws$_2$/graphene heterostructures grown by chemical vapor deposition}.
\newblock \emph{\bibinfo{journal}{Applied Physics Letters}}
  \textbf{\bibinfo{volume}{116}} (\bibinfo{year}{2020}).

\bibitem{zhu2013strain}
\bibinfo{author}{Zhu, C.} \emph{et~al.}
\newblock \bibinfo{title}{Strain tuning of optical emission energy and
  polarization in monolayer and bilayer mos$_2$}.
\newblock \emph{\bibinfo{journal}{Physical Review B}}
  \textbf{\bibinfo{volume}{88}}, \bibinfo{pages}{121301}
  (\bibinfo{year}{2013}).

\bibitem{feierabend2017impact}
\bibinfo{author}{Feierabend, M.}, \bibinfo{author}{Morlet, A.},
  \bibinfo{author}{Bergh{\"a}user, G.} \& \bibinfo{author}{Malic, E.}
\newblock \bibinfo{title}{Impact of strain on the optical fingerprint of
  monolayer transition-metal dichalcogenides}.
\newblock \emph{\bibinfo{journal}{Physical Review B}}
  \textbf{\bibinfo{volume}{96}}, \bibinfo{pages}{045425}
  (\bibinfo{year}{2017}).

\bibitem{kourmoulakis2023biaxial}
\bibinfo{author}{Kourmoulakis, G.} \emph{et~al.}
\newblock \bibinfo{title}{Biaxial strain tuning of exciton energy and
  polarization in monolayer ws$_2$}.
\newblock \emph{\bibinfo{journal}{Applied Physics Letters}}
  \textbf{\bibinfo{volume}{123}} (\bibinfo{year}{2023}).

\bibitem{an2023strain}
\bibinfo{author}{An, Z.} \emph{et~al.}
\newblock \bibinfo{title}{Strain control of exciton and trion spin-valley
  dynamics in monolayer transition metal dichalcogenides}.
\newblock \emph{\bibinfo{journal}{Physical Review B}}
  \textbf{\bibinfo{volume}{108}}, \bibinfo{pages}{L041404}
  (\bibinfo{year}{2023}).

\bibitem{zhang2022prolonging}
\bibinfo{author}{Zhang, Q.} \emph{et~al.}
\newblock \bibinfo{title}{Prolonging valley polarization lifetime through
  gate-controlled exciton-to-trion conversion in monolayer molybdenum
  ditelluride}.
\newblock \emph{\bibinfo{journal}{Nature Communications}}
  \textbf{\bibinfo{volume}{13}}, \bibinfo{pages}{4101} (\bibinfo{year}{2022}).

\bibitem{park2022efficient}
\bibinfo{author}{Park, S.} \emph{et~al.}
\newblock \bibinfo{title}{Efficient valley polarization of charged excitons and
  resident carriers in molybdenum disulfide monolayers by optical pumping}.
\newblock \emph{\bibinfo{journal}{Communications Physics}}
  \textbf{\bibinfo{volume}{5}}, \bibinfo{pages}{73} (\bibinfo{year}{2022}).

\bibitem{feng2019engineering}
\bibinfo{author}{Feng, S.} \emph{et~al.}
\newblock \bibinfo{title}{Engineering valley polarization of monolayer ws$_2$:
  a physical doping approach}.
\newblock \emph{\bibinfo{journal}{Small}} \textbf{\bibinfo{volume}{15}},
  \bibinfo{pages}{1805503} (\bibinfo{year}{2019}).

\bibitem{shinokita2019continuous}
\bibinfo{author}{Shinokita, K.} \emph{et~al.}
\newblock \bibinfo{title}{Continuous control and enhancement of excitonic
  valley polarization in monolayer wse$_2$ by electrostatic doping}.
\newblock \emph{\bibinfo{journal}{Advanced Functional Materials}}
  \textbf{\bibinfo{volume}{29}}, \bibinfo{pages}{1900260}
  (\bibinfo{year}{2019}).

\bibitem{katsipoulaki2023electron}
\bibinfo{author}{Katsipoulaki, E.} \emph{et~al.}
\newblock \bibinfo{title}{Electron density control in wse$_2$ monolayers via
  photochlorination}.
\newblock \emph{\bibinfo{journal}{2D Materials}} \textbf{\bibinfo{volume}{10}},
  \bibinfo{pages}{045008} (\bibinfo{year}{2023}).

\bibitem{robert2021measurement}
\bibinfo{author}{Robert, C.} \emph{et~al.}
\newblock \bibinfo{title}{Measurement of conduction and valence bands g-factors
  in a transition metal dichalcogenide monolayer}.
\newblock \emph{\bibinfo{journal}{Physical Review Letters}}
  \textbf{\bibinfo{volume}{126}}, \bibinfo{pages}{067403}
  (\bibinfo{year}{2021}).

\bibitem{shree2021guide}
\bibinfo{author}{Shree, S.}, \bibinfo{author}{Paradisanos, I.},
  \bibinfo{author}{Marie, X.}, \bibinfo{author}{Robert, C.} \&
  \bibinfo{author}{Urbaszek, B.}
\newblock \bibinfo{title}{Guide to optical spectroscopy of layered
  semiconductors}.
\newblock \emph{\bibinfo{journal}{Nature Reviews Physics}}
  \textbf{\bibinfo{volume}{3}}, \bibinfo{pages}{39--54} (\bibinfo{year}{2021}).

\bibitem{beret2022exciton}
\bibinfo{author}{Beret, D.} \emph{et~al.}
\newblock \bibinfo{title}{Exciton spectroscopy and unidirectional transport in
  mose$_2$-wse$_2$ lateral heterostructures encapsulated in hexagonal boron
  nitride}.
\newblock \emph{\bibinfo{journal}{npj 2D Materials and Applications}}
  \textbf{\bibinfo{volume}{6}}, \bibinfo{pages}{84} (\bibinfo{year}{2022}).

\bibitem{paradisanos2020controlling}
\bibinfo{author}{Paradisanos, I.} \emph{et~al.}
\newblock \bibinfo{title}{Controlling interlayer excitons in mos$_2$ layers
  grown by chemical vapor deposition}.
\newblock \emph{\bibinfo{journal}{Nature communications}}
  \textbf{\bibinfo{volume}{11}}, \bibinfo{pages}{2391} (\bibinfo{year}{2020}).

\bibitem{ren2023control}
\bibinfo{author}{Ren, L.} \emph{et~al.}
\newblock \bibinfo{title}{Control of the bright-dark exciton splitting using
  the lamb shift in a two-dimensional semiconductor}.
\newblock \emph{\bibinfo{journal}{Physical Review Letters}}
  \textbf{\bibinfo{volume}{131}}, \bibinfo{pages}{116901}
  (\bibinfo{year}{2023}).

\bibitem{demeridou2018spatially}
\bibinfo{author}{Demeridou, I.} \emph{et~al.}
\newblock \bibinfo{title}{Spatially selective reversible charge carrier density
  tuning in ws$_2$ monolayers via photochlorination}.
\newblock \emph{\bibinfo{journal}{2D Materials}} \textbf{\bibinfo{volume}{6}},
  \bibinfo{pages}{015003} (\bibinfo{year}{2018}).

\bibitem{sahin2013anomalous}
\bibinfo{author}{Sahin, H.} \emph{et~al.}
\newblock \bibinfo{title}{Anomalous raman spectra and thickness-dependent
  electronic properties of wse$_2$}.
\newblock \emph{\bibinfo{journal}{Physical Review B}}
  \textbf{\bibinfo{volume}{87}}, \bibinfo{pages}{165409}
  (\bibinfo{year}{2013}).

\bibitem{hanbicki2015measurement}
\bibinfo{author}{Hanbicki, A.}, \bibinfo{author}{Currie, M.},
  \bibinfo{author}{Kioseoglou, G.}, \bibinfo{author}{Friedman, A.} \&
  \bibinfo{author}{Jonker, B.}
\newblock \bibinfo{title}{Measurement of high exciton binding energy in the
  monolayer transition-metal dichalcogenides ws$_2$ and wse$_2$}.
\newblock \emph{\bibinfo{journal}{Solid State Communications}}
  \textbf{\bibinfo{volume}{203}}, \bibinfo{pages}{16--20}
  (\bibinfo{year}{2015}).

\bibitem{del2014excited}
\bibinfo{author}{Del~Corro, E.} \emph{et~al.}
\newblock \bibinfo{title}{Excited excitonic states in 1l, 2l, 3l, and bulk
  wse$_2$ observed by resonant raman spectroscopy}.
\newblock \emph{\bibinfo{journal}{Acs Nano}} \textbf{\bibinfo{volume}{8}},
  \bibinfo{pages}{9629--9635} (\bibinfo{year}{2014}).

\bibitem{yang2022relaxation}
\bibinfo{author}{Yang, M.} \emph{et~al.}
\newblock \bibinfo{title}{Relaxation and darkening of excitonic complexes in
  electrostatically doped monolayer wse$_2$: Roles of exciton-electron and
  trion-electron interactions}.
\newblock \emph{\bibinfo{journal}{Physical Review B}}
  \textbf{\bibinfo{volume}{105}}, \bibinfo{pages}{085302}
  (\bibinfo{year}{2022}).

\bibitem{courtade2017charged}
\bibinfo{author}{Courtade, E.} \emph{et~al.}
\newblock \bibinfo{title}{Charged excitons in monolayer wse$_2$: Experiment and
  theory}.
\newblock \emph{\bibinfo{journal}{Physical Review B}}
  \textbf{\bibinfo{volume}{96}}, \bibinfo{pages}{085302}
  (\bibinfo{year}{2017}).

\bibitem{he2020valley}
\bibinfo{author}{He, M.} \emph{et~al.}
\newblock \bibinfo{title}{Valley phonons and exciton complexes in a monolayer
  semiconductor}.
\newblock \emph{\bibinfo{journal}{Nature communications}}
  \textbf{\bibinfo{volume}{11}}, \bibinfo{pages}{618} (\bibinfo{year}{2020}).

\bibitem{ye2014probing}
\bibinfo{author}{Ye, Z.} \emph{et~al.}
\newblock \bibinfo{title}{Probing excitonic dark states in single-layer
  tungsten disulphide}.
\newblock \emph{\bibinfo{journal}{Nature}} \textbf{\bibinfo{volume}{513}},
  \bibinfo{pages}{214--218} (\bibinfo{year}{2014}).

\bibitem{kim1994thermodynamics}
\bibinfo{author}{Kim, J.}, \bibinfo{author}{Wake, D.} \&
  \bibinfo{author}{Wolfe, J.}
\newblock \bibinfo{title}{Thermodynamics of biexcitons in a gaas quantum well}.
\newblock \emph{\bibinfo{journal}{Physical Review B}}
  \textbf{\bibinfo{volume}{50}}, \bibinfo{pages}{15099} (\bibinfo{year}{1994}).

\bibitem{barbone2018charge}
\bibinfo{author}{Barbone, M.} \emph{et~al.}
\newblock \bibinfo{title}{Charge-tuneable biexciton complexes in monolayer
  wse$_2$}.
\newblock \emph{\bibinfo{journal}{Nature communications}}
  \textbf{\bibinfo{volume}{9}}, \bibinfo{pages}{3721} (\bibinfo{year}{2018}).

\bibitem{paradisanos2017room}
\bibinfo{author}{Paradisanos, I.} \emph{et~al.}
\newblock \bibinfo{title}{Room temperature observation of biexcitons in
  exfoliated ws$_2$ monolayers}.
\newblock \emph{\bibinfo{journal}{Applied Physics Letters}}
  \textbf{\bibinfo{volume}{110}} (\bibinfo{year}{2017}).

\bibitem{glazov2020optical}
\bibinfo{author}{Glazov, M.~M.}
\newblock \bibinfo{title}{Optical properties of charged excitons in
  two-dimensional semiconductors}.
\newblock \emph{\bibinfo{journal}{The Journal of Chemical Physics}}
  \textbf{\bibinfo{volume}{153}} (\bibinfo{year}{2020}).

\bibitem{maialle93}
\bibinfo{author}{Maialle, M.}, \bibinfo{author}{de~Andrada~e Silva, E.} \&
  \bibinfo{author}{Sham, L.}
\newblock \bibinfo{title}{Exciton spin dynamics in quantum wells}.
\newblock \emph{\bibinfo{journal}{Phys. Rev. B}} \textbf{\bibinfo{volume}{47}},
  \bibinfo{pages}{15776} (\bibinfo{year}{1993}).
\newblock \urlprefix\url{https://doi.org/10.1103/PhysRevB.47.15776}.

\bibitem{ivchenko05a}
\bibinfo{author}{Ivchenko, E.~L.}
\newblock \emph{\bibinfo{title}{Optical spectroscopy of semiconductor
  nanostructures}} (\bibinfo{publisher}{Alpha Science, Harrow UK},
  \bibinfo{year}{2005}).

\bibitem{glazov2014exciton}
\bibinfo{author}{Glazov, M.~M.} \emph{et~al.}
\newblock \bibinfo{title}{Exciton fine structure and spin decoherence in
  monolayers of transition metal dichalcogenides}.
\newblock \emph{\bibinfo{journal}{Physical Review B}}
  \textbf{\bibinfo{volume}{89}}, \bibinfo{pages}{201302}
  (\bibinfo{year}{2014}).

\bibitem{d1971spin}
\bibinfo{author}{D'yakonov, M.} \& \bibinfo{author}{Perel, V.}
\newblock \bibinfo{title}{Spin orientation of electrons associated with the
  interband absorption of light in semiconductors}.
\newblock \emph{\bibinfo{journal}{Soviet Journal of Experimental and
  Theoretical Physics}} \textbf{\bibinfo{volume}{33}}, \bibinfo{pages}{1053}
  (\bibinfo{year}{1971}).

\bibitem{konabe2016screening}
\bibinfo{author}{Konabe, S.}
\newblock \bibinfo{title}{Screening effects due to carrier doping on valley
  relaxation in transition metal dichalcogenide monolayers}.
\newblock \emph{\bibinfo{journal}{Applied Physics Letters}}
  \textbf{\bibinfo{volume}{109}} (\bibinfo{year}{2016}).

\bibitem{Kiselev74}
\bibinfo{author}{Kiselev, V.~A.} \& \bibinfo{author}{Zhilich, A.~G.}
\newblock \bibinfo{title}{Effective screening of the short-range exchange
  interaction in excitons}.
\newblock \emph{\bibinfo{journal}{Phys. Solid. State}}
  \textbf{\bibinfo{volume}{15}}, \bibinfo{pages}{1351} (\bibinfo{year}{1974}).

\bibitem{glazov2018breakdown}
\bibinfo{author}{Glazov, M.~M.} \& \bibinfo{author}{Chernikov, A.}
\newblock \bibinfo{title}{Breakdown of the static approximation for free
  carrier screening of excitons in monolayer semiconductors}.
\newblock \emph{\bibinfo{journal}{physica status solidi (b)}}
  \textbf{\bibinfo{volume}{255}}, \bibinfo{pages}{1800216}
  (\bibinfo{year}{2018}).

\bibitem{PhysRevB.102.125410}
\bibinfo{author}{Ayari, S.}, \bibinfo{author}{Jaziri, S.},
  \bibinfo{author}{Ferreira, R.} \& \bibinfo{author}{Bastard, G.}
\newblock \bibinfo{title}{Phonon-assisted exciton/trion conversion efficiency
  in transition metal dichalcogenides}.
\newblock \emph{\bibinfo{journal}{Phys. Rev. B}}
  \textbf{\bibinfo{volume}{102}}, \bibinfo{pages}{125410}
  (\bibinfo{year}{2020}).
\newblock \urlprefix\url{https://link.aps.org/doi/10.1103/PhysRevB.102.125410}.

\bibitem{beret2023nonlinear}
\bibinfo{author}{Beret, D.} \emph{et~al.}
\newblock \bibinfo{title}{Nonlinear diffusion of negatively charged excitons in
  monolayer wse$_2$}.
\newblock \emph{\bibinfo{journal}{Physical Review B}}
  \textbf{\bibinfo{volume}{107}}, \bibinfo{pages}{045420}
  (\bibinfo{year}{2023}).

\bibitem{astakhov00}
\bibinfo{author}{Astakhov, G.~V.} \emph{et~al.}
\newblock \bibinfo{title}{Oscillator strength of trion states in
  $\mbox{ZnSe}$-based quantum wells}.
\newblock \emph{\bibinfo{journal}{Phys. Rev. B}} \textbf{\bibinfo{volume}{62}},
  \bibinfo{pages}{10345} (\bibinfo{year}{2000}).

\bibitem{efimkin2017many}
\bibinfo{author}{Efimkin, D.~K.} \& \bibinfo{author}{MacDonald, A.~H.}
\newblock \bibinfo{title}{Many-body theory of trion absorption features in
  two-dimensional semiconductors}.
\newblock \emph{\bibinfo{journal}{Physical Review B}}
  \textbf{\bibinfo{volume}{95}}, \bibinfo{pages}{035417}
  (\bibinfo{year}{2017}).

\bibitem{wagner2023diffusion}
\bibinfo{author}{Wagner, K.} \emph{et~al.}
\newblock \bibinfo{title}{Diffusion of excitons in a two-dimensional fermi sea
  of free charges}.
\newblock \emph{\bibinfo{journal}{Nano Letters}} \textbf{\bibinfo{volume}{23}},
  \bibinfo{pages}{4708--4715} (\bibinfo{year}{2023}).

\bibitem{zhu2014exciton}
\bibinfo{author}{Zhu, C.} \emph{et~al.}
\newblock \bibinfo{title}{Exciton valley dynamics probed by kerr rotation in
  wse$_2$ monolayers}.
\newblock \emph{\bibinfo{journal}{Physical Review B}}
  \textbf{\bibinfo{volume}{90}}, \bibinfo{pages}{161302}
  (\bibinfo{year}{2014}).

\bibitem{prazdnichnykh2020control}
\bibinfo{author}{Prazdnichnykh, A.~I.} \emph{et~al.}
\newblock \bibinfo{title}{Control of the exciton valley dynamics in atomically
  thin semiconductors by tailoring the environment}.
\newblock \emph{\bibinfo{journal}{Phys. Rev. B}}
  \textbf{\bibinfo{volume}{103}}, \bibinfo{pages}{085302}
  (\bibinfo{year}{2021}).
\newblock \urlprefix\url{https://link.aps.org/doi/10.1103/PhysRevB.103.085302}.

\end{thebibliography}

\end{document}